\documentclass[%
reprint,
superscriptaddress,
 amsmath,amssymb,
 aps,
floatfix,
]{revtex4-1}
\usepackage{subfigure}
\usepackage{graphicx}% Include figure files
\usepackage{dcolumn}% Align table columns on decimal point
\usepackage{bm}% bold math
\usepackage[super]{nth}
\usepackage{mathtools}
\usepackage{lipsum}
\begin{document}

\title{Chiral nematic and fluctuation-induced first-order phase transitions in AB-stacked kagome bilayers}

\author{A. Zelenskiy}
 \affiliation{Department of Physics and Atmospheric Science, Dalhousie University, Halifax, Nova Scotia, Canada B3H 3J5}%Lines break automatically or can be forced with \\

\author{M. L. Plumer}
 \affiliation{Department of Physics and Atmospheric Science, Dalhousie University, Halifax, Nova Scotia, Canada B3H 3J5}
 \affiliation{Department of Physics and Physical Oceanography, Memorial University of Newfoundland, St. John’s, Newfoundland, A1B 3X7, Canada}
 
\author{B. W. Southern}
 \affiliation{Department of Physics and Astronomy, University of Manitoba, Winnipeg, Manitoba, Canada R3T 2N2}

\author{M. E. Zhitomirsky}
 \affiliation{Universit\'e Grenoble Alpes, Grenoble INP, CEA, IRIG, PHELIQS, 38000 Grenoble, France}
\affiliation{Institut Laue-Langevin, 71 Avenue des Martyrs, CS 20156, 38042 Grenoble Cedex 9, France}

\author{T. L. Monchesky}
\affiliation{Department of Physics and Atmospheric Science, Dalhousie University, Halifax, Nova Scotia, Canada B3H 3J5}%

\date{\today}% It is always \today, today,

\begin{abstract}
We study a Heisenberg-Dzyaloshinskii-Moriya Hamiltonian on AB-stacked kagome bilayers at finite temperature.
In a large portion of the parameter space, we observe three transitions upon cooling the system: a crossover from Heisenberg to the XY chiral paramagnet, Kosterlitz-Thouless transition to a chiral nematic phase, and a fluctuation-induced first-order transition to an Ising-like phase.
We characterize the properties of phases numerically using Monte Carlo finite-size analysis.
To further explain the nature of the observed phase transitions, we develop an analytical coarse-graining procedure that maps the Hamiltonian onto a generalized XY model on a triangular lattice.
To leading order, this effective model includes both bilinear and biquadratic interactions and is able to correctly predict the two phase transitions. 
Lastly, we study the Ising fluctuations at low temperatures and establish that the origin of the first-order transition stems from the quasi-degenerate ring manifold in the momentum space.
\end{abstract}

\pacs{Valid PACS appear here}% PACS, the Physics and Astronomy
\maketitle

%\tableofcontents

%%%%
%%%%
%Introduction
\textit{Introduction.}-- Competing interactions are at the root of complex behavior for a broad variety of physical systems~\cite{Muthukumar_Thomas_1997_science,Toledano_Zaccarelli_2009_sm,Ochi_Kuroki_2018_prb,Aka_Postma_2011_PLoS,Winter_2016_prb}.
In magnetic systems the competition can arise from lattice geometry (geometric frustration) or from the spin-orbit interactions~\cite{Savary_Balents_2017_rpp,Takagi_Nagler_2019_natrev}. 
The resulting ordered states, often possess important properties, such as topological stability and non-zero chirality, and are attractive for device applications~\cite{Fert_Sampaio_2013_nature}.
On the other hand, a considerable amount of research has been devoted to phases that either remain disordered down to zero temperature (spin liquids), or exhibit partial ordering, such as spin nematics~\cite{Savary_Balents_2017_rpp,Knolle_Moessner_2019_arcmp,penc_2010}.
The experimental discovery of these phases remains challenging, since they lack conventional dipolar ordering.
Theoretical studies have shown that the stabilization of spin nematic states is non-trivial~\cite{Jiang_Chernyshev_2023_prb} and often requires higher-order spin interactions, such as the biquadratic exchange~\cite{Blume_Hsieh_2003_jap,andreev_Grishchuk_1984_JETP,Shannon_Motome_2010_prb}.
Furthermore, since spin nematics break rotational, but not the time-reversal symmetry, the orientations of the spins in these phases continue to fluctuate in the Ising-like fashion.
Geometric frustration, among other things, was shown to accommodate for these fluctuations~\cite{Shannon_Sindzingre_2006_prl,Momoi_Kubo_2012_prl,Iqbal_Thomale_2016_prb}.

The Heisenberg antiferromagnet on a kagome lattice is the paradigmatic example of geometrically frustrated spin system with macroscopic degeneracy~\cite{Chalker_Holdsworth_Shender_kagome_1992_prl,Harris_Kallin_Berlinsky_kagome_1992_prb,Huse_Rutenberg_kagome_1992_prb}.
At low temperatures, the classical spins form the 120 coplanar structure with dominant octupolar correlations~\cite{Zhitomirsky_2008_prb}.
\begin{figure}[!th]
    \centering
    \includegraphics[width=0.5\textwidth]{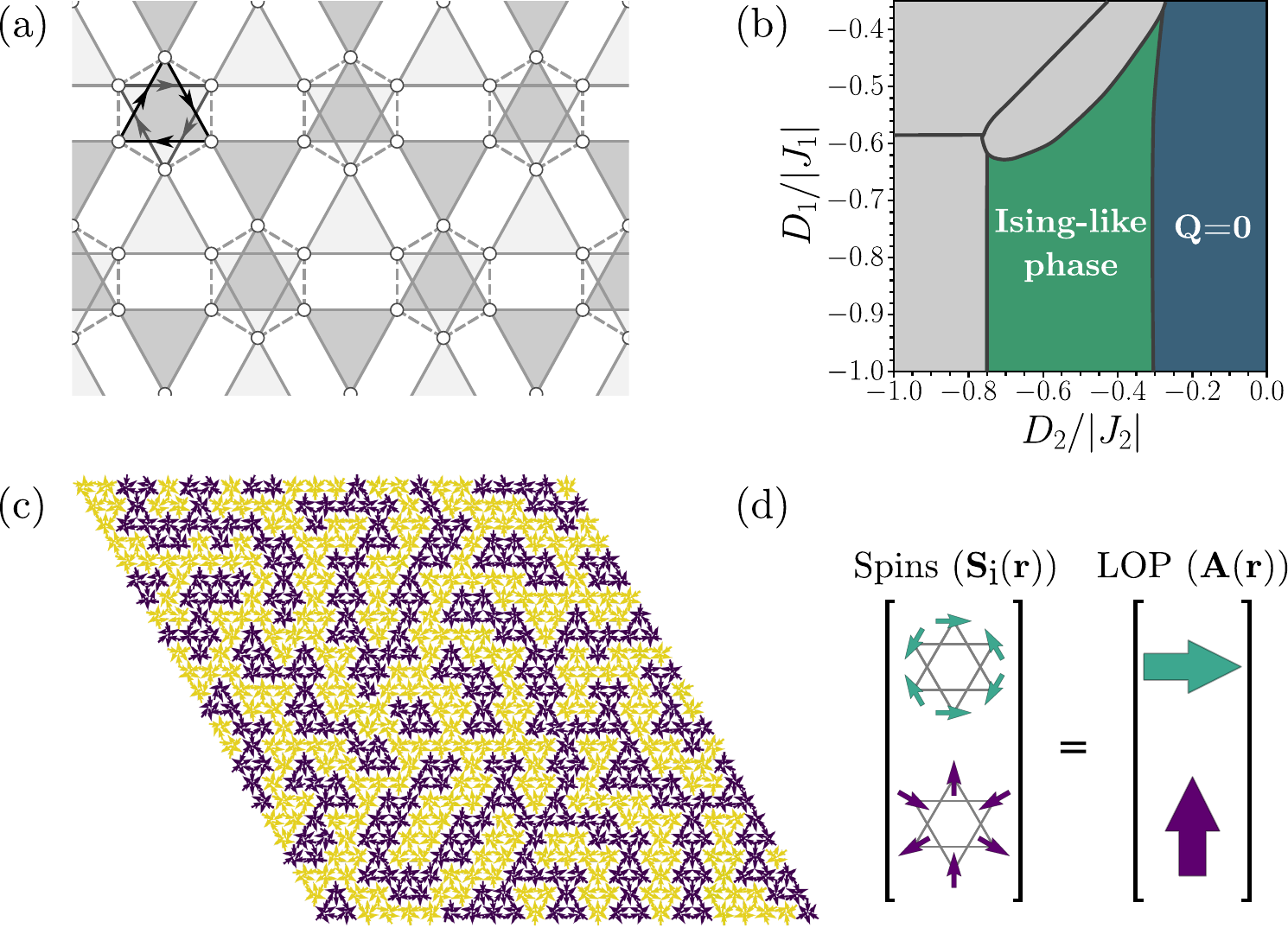}
    \caption{(a)~Crystal structure of the AB-stacked kagome compounds: the A and B layers are shown in two shades of gray, dashed and solid lines represent the ($J_1$, $D_1$) and ($J_2$, $D_2$) interactions respectively. The arrows on the top left cell indicate bond directions taken for calculating chirality $\chi(\mathbf{r})$. (b)~Part of the phase diagram (from Ref.~I) where the Ising-like phases are stable. (c)~Fragment of the spin structure in the Ising-like phase. The color corresponds to $\mathbf{A}(\mathbf{r})\cdot\mathbf{A}(0)$ (purple and yellow for $\pm 1$ respectively). (d)~Definition of the LOPs.}
    \label{fig:structure}
\end{figure}
The macroscopic degeneracy in 2D kagome is generally unstable with respect to anisotropic interactions, such as Dzyaloshinskii-Moriya (DM) or changes in the geometry that introduce competing interactions~\cite{Chernyshev_Zhitomirsky_2015_prb,McCoombs_Monchesky_2023_prb}.
\begin{figure*}[!ht]
    \centering
    \includegraphics[width=0.95\textwidth]{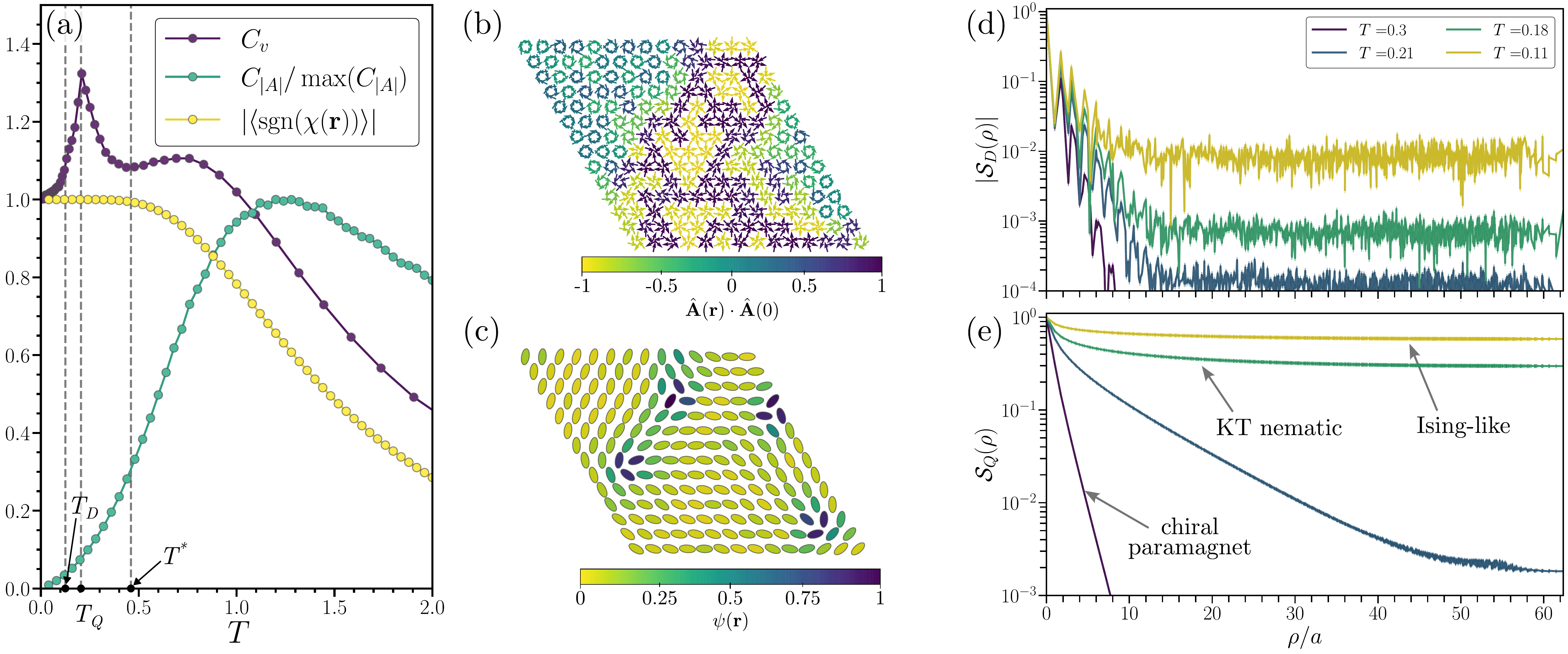}
    \caption{(a)~High-temperature properties of the heat-capacity, LOP magnitude susceptibility, and average chirality alignment. (b), (c)~Fragment of the spin configuration in the chiral nematic phase. In (b), we plot the spins on AB-SKL and color the unit cells using the overlap of the LOP vectors. In (c), we plot the same configuration but in terms of $\mathbf{A}(\mathbf{r})$, where the collinearity parameter (see text) is used to color the directors. The dark spots correspond to disclinations. (d)~Dipolar and (e)~quadrupolar correlation functions plotted as function of distance per kagome bond-length.}
    \label{fig:cv_chir_corr}
\end{figure*}

A previous study of the \textit{AB-stacked} kagome lattice (AB-SKL)~\cite{Zelenskiy_Monchesky_Plumer_Southern_2022_prb} (Ref.~I) revealed that the symmetry of the model introduces a large number of duality transformations, allowing for a unified description of magnetic phases in different parts of the parameter space. 
Furthermore, a minimal Heisenberg-DM (HDM) Hamiltonian was shown to stabilize various single- and \mbox{multiple-$q$} structures.
Among these, the most intriguing are magnetic phases where the spins in individual unit cells have a distorted 120 degree structure, which alternates throughout the system, forming Ising-like structures (see fig.~\ref{fig:structure}~(c)).

In this article, we study the finite-temperature properties of the Ising-like phase in a single AB-SKL bilayer.
We show through both numerical and analytical calculations that in a large region of parameter space thermal fluctuations stabilize a phase that exhibits simultaneous chiral and nematic order.
The coexistence of nematicity and chirality is extremely unusual, since these properties are typically associated with opposite types of structures (collinear and non-collinear, respectively).
Moreover, we find that at lower temperatures, the chiral nematic phase breaks the time-reversal symmetry and transforms into Ising-like structures via a fluctuation-induced first-order transition. 

Our study is relevant to the magnetic properties of compounds with AB-SKL structure, such as $\mathrm{Mn}_3X$ ($X = \mathrm{Sn}$, $\mathrm{Ge}$, $\mathrm{Ga}$) and $\mathrm{Fe}_3\mathrm{Sn}_2$.
These systems have received a considerable amount of attention due to their unusual transport properties, including recent discoveries of the anomalous Hall effect~\cite{Nakatsuji_Kiyohara_Higo_ahe_2015_nature,Nayak_ahe_2016_science,Kiyohara_ahe_2016_prap,Kida_Wills_2011_jpcm}.
The non-magnetic atoms in these materials induce a weak spin-orbit coupling, which by symmetry should result in the intra- and interlayer DM interactions~\cite{Zelenskiy_Monchesky_Plumer_Southern_2021_prb}.
Despite the experimental reports of helical~\cite{Park_gs_2018_nature_pub,Fenner_Wills_2009_jpcm}, skyrmion~\cite{Hirschberger_Gd_2019_nature_comm}, magnetic bubble, and spin glass~\cite{Hou_Zhang_2017_am} phases in AB-SKL materials, a theoretical description of the magnetism of these systems is still lacking.

%%%%
%%%%
%Model

\textit{Model Hamiltonian.}-- We consider classical $\mathrm{O}(3)$ spins on an AB-SKL.
The minimal HDM model was derived in~\cite{Zelenskiy_Monchesky_Plumer_Southern_2021_prb} and can be written as follows:

\begin{align}
    &\mathcal{H}_{JD} = \mathcal{H}_J + \mathcal{H}_D,\label{eq:JD_Hamiltonian}\\
    &\mathcal{H}_J = \frac{1}{2}\sum_{\mathbf{r}\mathbf{r'}} \sum_{ij} J_{ij}(\mathbf{r}-\mathbf{r}')\mathbf{S}_i(\mathbf{r})\cdot\mathbf{S}_j(\mathbf{r}'),\notag\\
    &\mathcal{H}_D = \frac{1}{2}\sum_{\mathbf{r}\mathbf{r'}} \sum_{ij} D_{ij}(\mathbf{r}-\mathbf{r}')\mathbf{\hat{z}}\cdot\left[\mathbf{S}_i(\mathbf{r})\times\mathbf{S}_j(\mathbf{r}')\right],\notag
\end{align} 
where $\mathbf{r}$ and $\mathbf{r}'$ label the positions of the unit cells in a triangular superlattice, and $i,j$ label the six sublattices.
The four parameters correspond to the interlayer ($J_1$, $D_1$) and intralayer ($J_2$, $D_2$) exchange and DM couplings, respectively (see fig.~\ref{fig:structure}~(a)). 
The interlayer interactions stabilize $Q_z=0$ for all relevant parameter values, and so the properties of a single bilayer should be representative of the properties in the bulk.  
In the following, we will also take advantage of the self-duality, by defining local coordinates for the six sublattices.
\begin{figure}[!ht]
    \centering
    \includegraphics[width=0.47\textwidth]{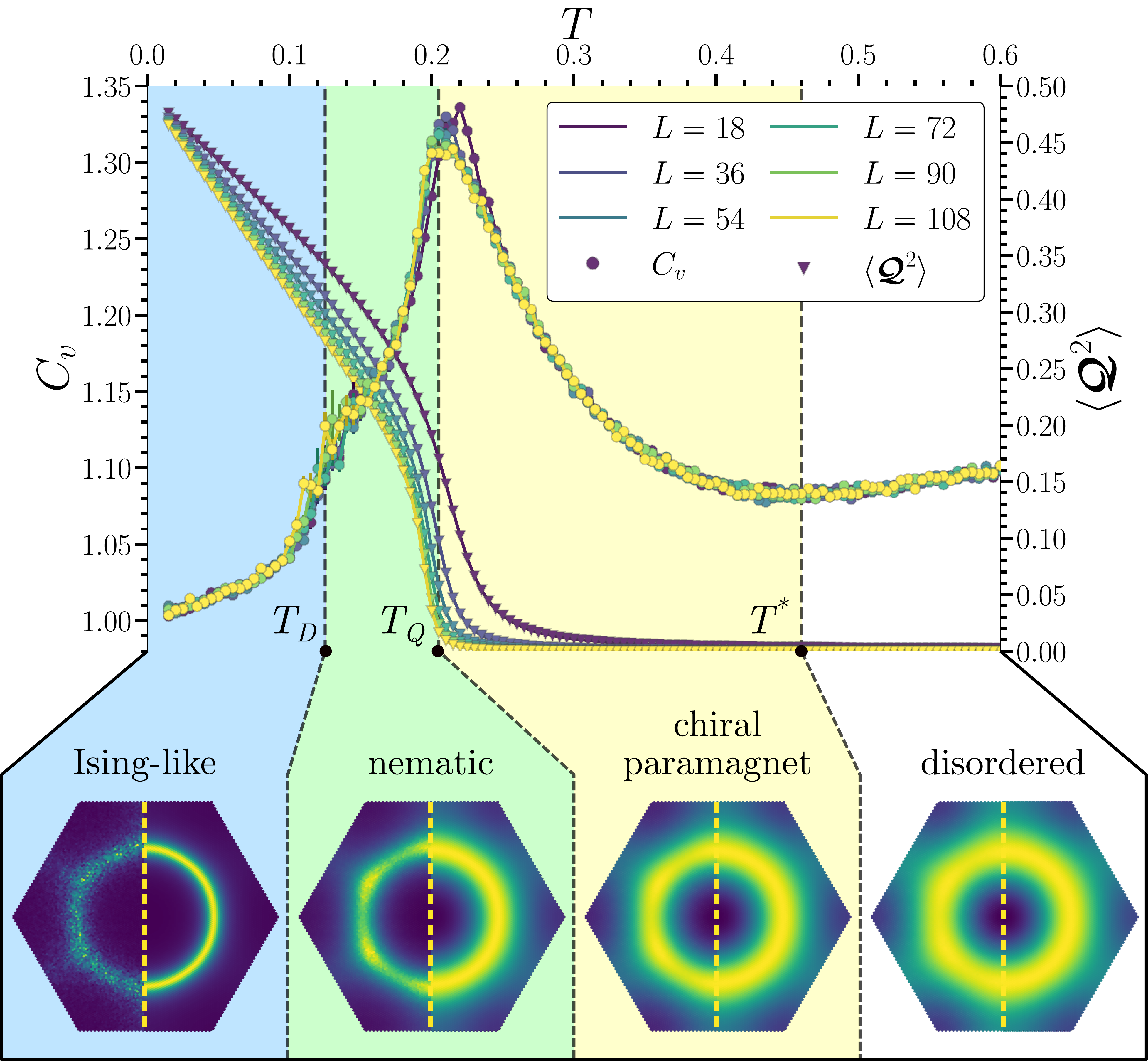}
    \caption{Magnetic phases studied in this work. (Top)~Finite-size data for low-temperature heat capacity and average quadrupole moment. (Bottom)~Average dipolar structure factors from MC data (left half) and calculated from the effective XY Hamiltonian (right half). }
    \label{fig:cv_quad_corr}
\end{figure}
The dual version of the model~(\ref{eq:JD_Hamiltonian}) is written in terms of new local spin variables $\tilde{\mathbf{S}}_i(\mathbf{r})$ as well as dual parameters  ($\widetilde{J}_1$, $\widetilde{D}_1$, $\widetilde{J}_2$, $\widetilde{D}_2$)~\cite{Supplemental}.
An important property of the dual parameters, pointed out in Ref.~I, is that in the stability region of the Ising-like phases (fig.~\ref{fig:structure}~(b)) we generally have $|\widetilde{D}_2|/|\widetilde{J}_1|< 1$ and $|\widetilde{J}_2|/|\widetilde{J}_1|\ll 1$.

%%%%
%%%%
%Simulation details
\textit{Details of numerical simulations.}-- We perform Monte Carlo (MC) simulations using standard heat-bath updates combined with the over-relaxation method~\cite{Creutz_1987_prd}.
A single MC step consists of one heat-bath update, followed by 5 over-relaxation steps.
Simulations are performed in bilayer systems with $N=L^2$ unit cells with $18\leq L \leq 108$, and in the temperature range $0.01\leq T \leq 5$.
A single run typically consists of $10^5$ MC steps at each temperature.
Finally, the results are averaged over 10 independent simulations to estimate the statistical errors.
A list of definitions of average quantities is provided in the Supplemental Material~\cite{Supplemental}.

For consistency with the results for the 3D systems in Ref.~I, we fix $J_1=2$ and $J_2=1$ and vary the values of $D_1$ and $D_2$.
Thus, the temperatures can be assumed to have units of $|J_2|$. 
In this article, we present the results for two representative systems with $D_1=-J_1$ and $D_2=-0.5J_2$ and provide data for the extended range of parameters in~\cite{Supplemental}.

%%%%
%%%%
%Monte Carlo results

%%%%
%%%%
%Chiral paramagnet

\textit{Monte Carlo results.}-- First, we report our numerical findings.
Since the DM term in our model~(\ref{eq:JD_Hamiltonian}) breaks the out-of-plane $C_2$ spin symmetry, the system develops non-zero chirality in each unit cell, which we define for the bilayer system as $\chi(\mathbf{r}) = \hat{\mathbf{z}}\cdot\sum_{\langle ij\rangle}\mathbf{S}_i(\mathbf{r})\times\mathbf{S}_j(\mathbf{r})$, where indices $i,j$ label sites on a kagome triangle (see fig.~\ref{fig:structure}).
As the system is cooled, qualitative changes in the spin structure occur at temperatures $T^*$, $T_Q$, $T_D$ (fig.~\ref{fig:cv_chir_corr}~(a)).
Above $T^*$, we observe a broad Schottky-like peak in the heat capacity.
A closer analysis reveals that in this temperature range the spins in each unit cell form an approximately 120 degree planar structure, with spins on the A triangle parallel to those on the B triangle (fig.~\ref{fig:structure}).
The corresponding ``local'' order parameter (LOP) $\mathbf{A}(\mathbf{r})$ is a two-dimensional vector, which transforms as an irreducible representation $E_g^{(14)}$, as discussed in Ref.~I.
Below $T^*$, the chirality in each unit cell becomes negative, \textit{i.e.} $\langle\mathrm{sgn}(\chi(\mathbf{r}))\rangle = -1$.
At the same time, the fluctuations in magnitude of LOPs become very small, as seen from the temperature dependence of $C_{|A|} = \frac{1}{N}\sum_\mathbf{r}\langle (|A(\mathbf{r})|^2-\langle |A(\mathbf{r})|^2\rangle)^2 \rangle$ in fig.~\ref{fig:cv_chir_corr}~(a).

Despite the apparent ordering of the chiralities, the in-plane spin fluctuations remain large, which poses a question about the global ordering of the system.
To analyze the spin structure on the global scale, we define the dipole and quadrupole correlation functions respectively as 
\begin{align}
    \mathcal{S}_D(\boldsymbol{\rho}) &= \frac{1}{N}\sum_{\mathbf{r}}\langle \hat{\mathbf{A}}(\mathbf{r})\cdot\hat{\mathbf{A}}(\mathbf{r}+\boldsymbol{\rho})\rangle,\\
    \mathcal{S}_Q(\boldsymbol{\rho}) &= \frac{1}{N}\sum_{\mathbf{r}}\langle \boldsymbol{\mathcal{Q}}(\mathbf{r})\cdot\boldsymbol{\mathcal{Q}}(\mathbf{r}+\boldsymbol{\rho})\rangle,
\end{align}
where ${\boldsymbol{\rho}=\mathbf{r}-\mathbf{r}'}$, and the quadrupole tensor is defined as ${\mathcal{Q}_{\alpha\beta}(\mathbf{r})=\hat{A}_\alpha(\mathbf{r})\hat{A}_\beta(\mathbf{r})-\frac{1}{2}\delta_{\alpha\beta}}$.
As seen from fig.~\ref{fig:cv_chir_corr}~(d),~(e), in the temperature range between $T_Q$ and $T^*$, both types of correlations decay exponentially with distance.
Thus, in this region, the state of the system can be thought of as \textit{chiral paramagnet}~\cite{Pappas_Farago_2009_prl}.

%%%%
%%%%
%Nematic transition
Further decreasing the temperature of the system, we observe an appearance of spontaneous quadrupole moment at $T_Q$ (fig.~\ref{fig:cv_quad_corr}).
Below $T_Q$, $\mathcal{S}_Q(\rho=|\boldsymbol{\rho}|)$ displays a clear algebraic decay, with a correlation length that strongly depends on the temperature (fig.~\ref{fig:cv_chir_corr}~(e)).
This is a strong indication of the emergent quasi-long-range KT ordering of the nematic degrees of freedom (NDOF), as a result of the algebraic breaking of the continuous $U(1)$ symmetry.
Since chirality vectors remain ordered, this nematic phase is also chiral.
We confirm the KT nematic order by defining the collinearity parameter ${\psi(\mathbf{r}) = \frac{1}{3}\sum_{\boldsymbol{\rho}} \langle\boldsymbol{\mathcal{Q}}(\mathbf{r})\cdot\boldsymbol{\mathcal{Q}}(\mathbf{r}+\boldsymbol{\rho})\rangle}$, where the sum runs over the nearest neighbors~\cite{Supplemental}.
Fig.~\ref{fig:cv_chir_corr}~(c) and~\cite{Supplemental} shows the formation of topological defects, identical in character to the disclinations in conventional nematics~\cite{deGennes_1993}.
Below $T_Q$, the defects appear in pairs, which is further consistent with the KT theory of 2D nematics~\cite{Stein_1978_prb}.

In the reciprocal space, $\mathcal{S}_Q(\mathbf{q})$ becomes sharply peaked at $\mathbf{q}=0$ at $T_Q$~\cite{Supplemental}.
In contrast, $\mathcal{S}_D(\mathbf{q})$ shows a broad ring feature at incommensurate wavevectors~(fig.~\ref{fig:cv_quad_corr}).
As the system is cooled down below $T_Q$, the number of wavevectors contributing to the ring decreases and the six-fold anisotropy becomes more pronounced.
Since the Ising degrees of freedom (IDOF) continue to fluctuate, we conclude that the time-reversal $\mathbb{Z}_2$ symmetry must remain unbroken in the nematic phase.

%%%%
%%%%
%Brazovskii transition
\begin{figure}[!ht]
    \centering
    \includegraphics[width=0.48\textwidth]{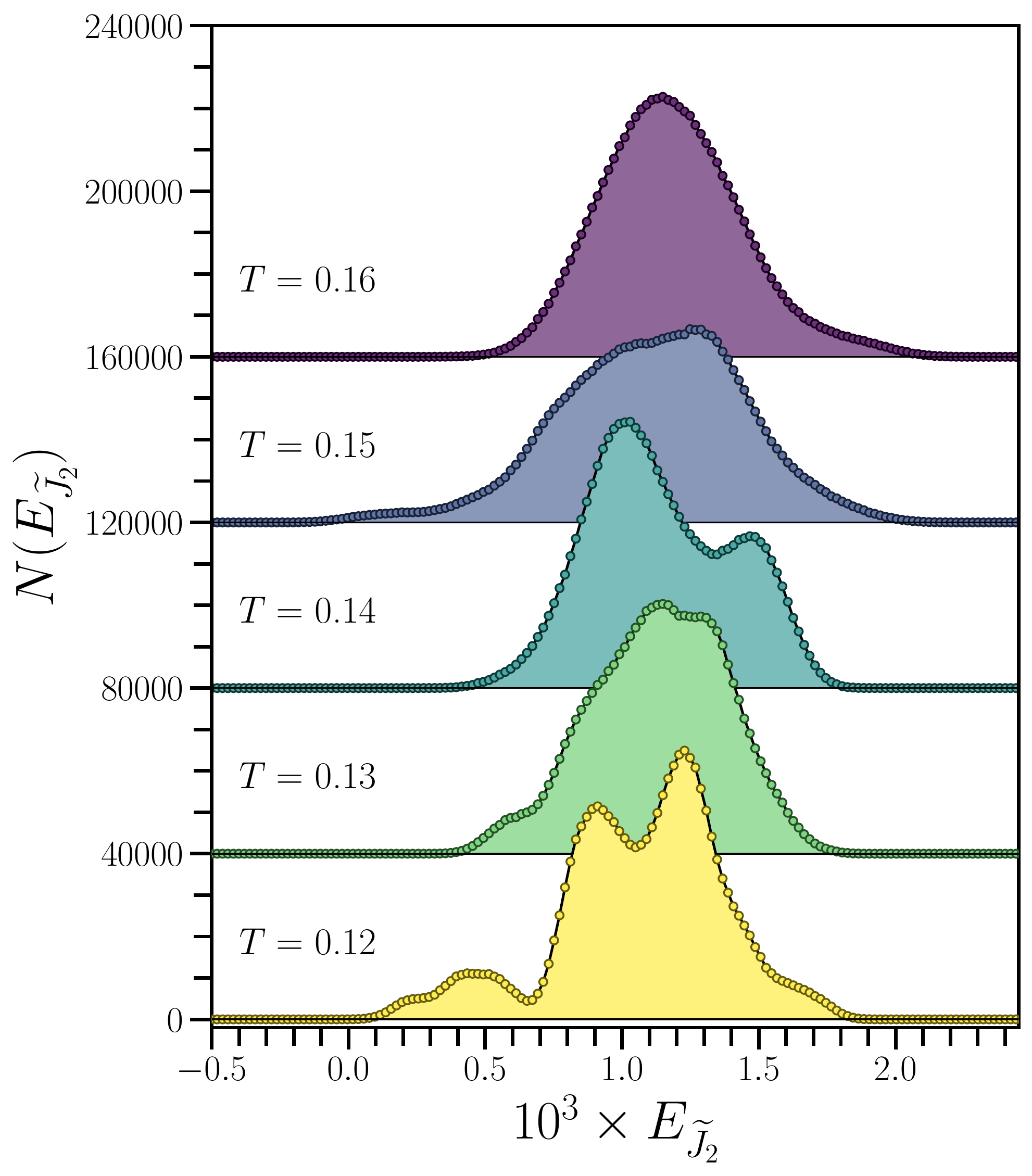}
    \caption{Energy histograms for a range of temperatures near $T=T_D$ for $L=54$ system.}
    \label{fig:histograms}
\end{figure}
Finally, at $T_D$ the IDOF freeze, and the LOP vectors form complicated network patterns, as reported in Ref.~I.
Analysis of the histograms of energy components at $T_D$ (collected using $2\times10^6$ MC steps) reveal multiple peaks (fig.~\ref{fig:histograms}), which signals a weak first-order transition.
Unlike in the conventional first-order transitions, we observe 3-7 peaks in the energy histograms for a range of temperatures. 
The heat capacity in the same range of temperatures appears noisy and does not display a clear anomaly. 

These observations lead us to believe that the free energy landscape of the Ising-like phases consists of many near-degenerate minima. 
Therefore, we suspect that the characteristic network patterns form when the system fails to reach equilibrium as a result of the large configurational entropy. 
Our speculations are supported by the fact that for most systems in our study we can impose a \mbox{single-$q$} stripe structure which is slightly lower in energy than the random configurations ($\Delta E\sim 10^{-5}$). 
Nevertheless, these ordered states almost never occur in 2D system, even for longer MC runs ($> 10^6$ MC updates). 

%%%%
%%%%
%Coarse-graining
\textit{Coarse-graining.}-- To better understand the MC data, we provide analytical analysis of the model~(\ref{eq:JD_Hamiltonian}).
When the system is in the chiral paramagnet state, its properties are effectively described by $N$ fluctuating LOP vectors.
Therefore, it is desirable to construct an effective Hamiltonian, written explicitly in terms of $\mathbf{A}(\mathbf{r})$ variables. 
To do this, we consider the normal modes of spin fluctuations within a single unit cell~\cite{Dasgupta_Tchernyshyov_2020_prb}.
In total, there are twelve modes, half of which ($\alpha_{\{0-5\}}(\mathbf{r})$) describe the in-plane fluctuations, and the remaining ($\gamma_{\{0-5\}}(\mathbf{r})$) describe the out-of-plane fluctuations.
Among these, only the uniform in-plane rotations, which we denote $\alpha_0(\mathbf{r})$, do not change the magnitude of the LOP vectors.
As a result, we may construct a coarse-graining procedure, whereby the hard modes $\alpha_{\{1-5\}}(\mathbf{r})$, $\gamma_{\{0-5\}}(\mathbf{r})$ are integrated out, leaving effective interactions written in terms of the soft modes $\alpha_0(\mathbf{r})$.
The procedure follows closely the method presented in Ref.~I, and is given in~\cite{Supplemental}.
In the derivation, we take advantage of the smallness of $|\widetilde{J_2}|/|\widetilde{J}_1|\sim 10^{-2}$ and calculate the effective Hamiltonaian up to the smallest power of $|\widetilde{D_2}|/|\widetilde{J}_1|\sim 10^{-1}$.
The result is a generalized XY Hamiltonian on a triangular lattice:
\begin{align}
    \mathcal{H}_\text{eff} &= E_0 + \mathcal{H}_D + \mathcal{H}_Q + \mathcal{H}_{DQ},\label{eq:eff_ham}\\
    \mathcal{H}_D &= \frac{1}{2}\sum_{\mathbf{r}\mathbf{r}'} \mathcal{J}_D(\boldsymbol{\rho})\hat{\mathbf{A}}(\mathbf{r})\cdot \hat{\mathbf{A}}(\mathbf{r}'),\label{eq:eff_ham_D}\\
    \mathcal{H}_Q &= \frac{1}{2}\sum_{\mathbf{r}\mathbf{r}'} \mathcal{J}_Q(\boldsymbol{\rho})\boldsymbol{\mathcal{Q}}(\mathbf{r})\cdot \boldsymbol{\mathcal{Q}}(\mathbf{r}'),\label{eq:eff_ham_Q}\\
    \mathcal{H}_{DQ} &= \frac{1}{2}\sum_{\mathbf{r}\mathbf{r}'\mathbf{r}''} \mathcal{J}_{DQ}(\boldsymbol{\rho};\boldsymbol{\rho}')\hat{\mathbf{A}}^T(\mathbf{r}')\boldsymbol{\mathcal{Q}}(\mathbf{r}) \hat{\mathbf{A}}(\mathbf{r}''),\label{eq:eff_ham_DQ}
\end{align}
where $E_0$ is a constant, ${\boldsymbol{\rho}=\mathbf{r}-\mathbf{r}'}$, and ${\boldsymbol{\rho}'=\mathbf{r}-\mathbf{r}''}$.
Importantly, $\mathcal{H}_Q$ is equivalent to a biquadratic coupling of the LOP vectors~\cite{Supplemental}. 
The dipolar couplings $\mathcal{J}_D(\boldsymbol{\rho})$ extend to the third neighbors and lead to geometric frustration, whereas $\mathcal{J}_Q(\boldsymbol{\rho})$ only couple nearest neighbors and stabilize collinear (nematic) configurations of the LOPs.
Generally, the biquadratic couplings are larger than the dipolar (${|\mathcal{J}_Q(\boldsymbol{\rho})|\sim~5|\mathcal{J}_D(\boldsymbol{\rho})|}$).

A family of similar generalized XY models has been studied numerically~\cite{Granato_Nightingale_1991_prb,Carpenter_Chalker_1989_jpcm,Poderoso_Levin_2011_prl,Shi_Fendley_2011_prl,Canova_Arenson_2014_pre,Canova_Arenson_2016_pre,Serna_Fendley_2017_jpamt,Zukovic_2018_pla} and using renormalization techniques~\cite{Lee_Grinstein_1985_prl,Korshunov_1985_JETP,Korshunov_1986_jpcssp}.
In these works, the dipolar interactions are typically unfrustrated and stabilize ferromagnetic order.
When the coupling term is zero, the phase diagram in the $|\mathcal{J}_D(\boldsymbol{\rho})|/|\mathcal{J}_Q(\boldsymbol{\rho})|\ll 1$ limit has been well established: as the system is cooled down, it first undergoes a nematic KT transition ($T=T_Q$), followed by an Ising transition ($T=T_D$) leading to a phase with a quasi-long-range ferromagnetic order.
This model was proposed to be relevant for a range of systems, including liquid crystals, and superconductors~\cite{Lee_Grinstein_1985_prl,Korshunov_1985_JETP,Shi_Fendley_2011_prl,Serna_Fendley_2017_jpamt}.
In the vast majority of magnetic systems, the biquadratic term ($\mathcal{H}_Q$), if present, is smaller than the exchange interaction, meaning that the split transition cannot occur through this mechanism.
Our coarse-graining procedure uncovers that the effective interactions impose a large quadrupolar coupling through DM interactions.
We note that the validity of the effective model in~(\ref{eq:eff_ham}) extends beyond the Ising-like phases into the $\mathbf{Q}=0$ phase. 
Using duality transformations in Ref.~I, we can quickly construct similar models for other $\mathbf{Q}=0$ phases. 
Since these phases occupy most of the parameter space, and are known to be the ground states of the $\mathrm{Mn}_3X$ compounds, the properties of~(\ref{eq:eff_ham}) are extremely relevant for the future experimental studies.

In the case of the Ising-like phases, the situation is complicated by both the geometric frustration on the dipolar interaction and the presence of the coupling term. 
 $\mathcal{J}_{D}(\mathbf{q})$ produces a degenerate ring, similar to the $\mathcal{S}_D(\mathbf{q})$, which leads to a competition between different incommensurate configurations.
The $\mathcal{H}_{DQ}$ may influence a variety of properties, including the universality classes of transitions, as well as the nature of topological defects~\cite{Granato_Nightingale_1991_prb,Drouin-Touchette_Lubensky_2022_prx,Jiang_Huang_1993_pre,Jiang_Huang_1996_prl}.
This analysis is outside the scope of this work and will be reported elsewhere~\cite{unpublished}
Our calculations show that $\mathcal{H}_{DQ}$ changes the value of $T_D$ as well as the radius of the ring in $\mathcal{J}_D(\mathbf{q})$, but does not break its degeneracy~\cite{Supplemental}. 
Therefore, our results for the decoupled model ($\mathcal{H}_{DQ}=0$) still apply to the physics of the system.

%%%%
%%%%
%Mean-field theory
\textit{Mean-field theory.}-- In order to study the properties of the effective model in eq.~(\ref{eq:eff_ham}), we construct a mean-field theory using variational methods~\cite{Reimers_Shi_1991_prb,Berlinsky_Harris_2019}.
The derivation of the model is given in the Supplemental Material~\cite{Supplemental}.
We denote $Q$ and $\phi(\mathbf{q})$ as the order parameters for the NDOF and the IDOF, respectively.
We obtain the following Landau expansion

\begin{align}
    f_L &= f_0 + f_D + f_Q,\label{eq:Landau_model}\\
    f_D &= \tau_D \Phi + 3\lambda_D\Phi^2 - \frac{3\lambda_D}{2}\sum_\mathbf{q}|\phi(\mathbf{q})|^4,\label{eq:Landau_D}\\
    f_Q &= \frac{\tau_Q}{2} Q^2 + \frac{\lambda_Q}{4} Q^4,\label{eq:Landau_Q}
\end{align}
where $f_0$ is a constant, and $\Phi = \sum_\mathbf{q} |\phi(\mathbf{q})|^2$.
Here, we restrict the wavevectors to lie on the degenerate ring.
The coefficients $\tau_D,\tau_Q$ change sign at $T_D$ and $T_Q$ respectively, and are related to the corresponding bare susceptibilities evaluated at the critical wavevectors, and $\lambda_D, \lambda_Q$ are positive constants~\cite{Supplemental}.

The mean-field theory predicts two phase transitions, consistent with the numerical results.
We note that $f_Q$ has exactly the same form as the mean-field expansion for a 2D XY model, which is unsurprising given that $\mathcal{H}_Q$ can be mapped onto an XY Hamiltonian by changing $\alpha_0(\mathbf{r})\longrightarrow \frac{1}{2}\alpha_0(\mathbf{r})$.
Therefore, in the decoupled limit, the nematic transition should belong to the XY universality class.
This is in contrast to a system with a 3D nematic order parameter, where a first-order transition is predicted at the level of the mean-field theory.

The model further predicts the Ising transition to be continuous.
Assuming that the order parameter is defined by $m$ magnetic wavevectors, the corresponding free energy in the ordered state is 

\begin{equation}
    f_D^\text{min} = -\frac{\tau_D^2}{6\lambda_D}\frac{m}{2m-1}.
\end{equation}
The structure of the equilibrium Ising order parameter depends on the sign of $\lambda_D$.
Thus, in the mean-field limit, the free energy is minimized by a \mbox{single-$q$} solution ($m~=~1$).

%%%%
%%%%
%Fluctuations
\textit{Effects of fluctuations.}-- This predicted nature of the phase transition is inconsistent with our numerical observations, which indicates that the role of thermal fluctuations is not negligible.
Indeed, the degenerate ring of critical wavevectors $\mathcal{H}_D$ signals that the phase space of fluctuations is very large, even if their amplitudes are small.
Near $T=T_D$, the bare dipolar susceptibility can be parameterized according to 

\begin{equation}
    \chi_{0,D}(\mathbf{q}) \approx \frac{1}{\tau_D + c(q-q_0)^2},
\end{equation}
where we ignored the effects of the hexagonal anisotropy.
We further consider only wavevectors with radius $q$ close to the critical ring ($q_0$).

This scenario was first studied in three-dimensional isotropic systems by Brazovskii~\cite{Brazovskii_1975_JETP}, who, for the case of a 3D system, showed that the large volume of fluctuations stabilizes the disordered state down to $T=0$.
This prevents the system from undergoing a continuous phase transition.
Nevertheless, Brazovskii's analysis indicated that the system may still have a first-order transition, even in the absence of a cubic term in the Landau theory.

Since the field theory for the IDOF in the AB-SKL is identical to that contained in Ref~\cite{Brazovskii_1975_JETP}, up to the dimension of the system, we follow the same steps to obtain the renormalized values of $\tau_D$ and $\lambda_D$~\cite{Chaikin_Lubensky_2000,Altland_Simons_2010,Amit_2005}.
The procedure is described in the Supplemental Material~\cite{Supplemental}.
For the renormalized susceptibility, we obtain

\begin{equation}
    \chi_D(\mathbf{q}) = \frac{1}{t_D + c(q-q_0)^2},
\end{equation}
where the renormalized parameter $t_D$ is defined through a simple self-consistency relation:

\begin{equation}
    t_D= \tau_D + \frac{3\lambda_Dq_0}{2\sqrt{ct_D}}.
\end{equation}
Since $t_D$ is non-negative for all values of $\tau_D$, the fluctuations stabilize the nematic phase for all $T<T_Q$.
Furthermore, the renormalized value of the vertex $\lambda_D$ is calculated to be

\begin{equation}
    l_D = \lambda_D\frac{1-2\Pi}{1+\Pi},
\end{equation}
where $\Pi\propto t_D^{-\frac{3}{2}}$.
Since $l_D$ changes sign at $2\Pi=1$, and since mean-field theory predicts a positive sixth-order term~\cite{Supplemental}, we conclude that thermal fluctuations induce a first-order transition.
The negative value of $l_D$ and the form of eq.~(12) further indicate that fluctuations will prefer \mbox{multiple-$q$} solutions, which can contribute to the formation of random Ising-like structures.

The Brazovskii transition has been mostly discussed in the context of weak crystallization~\cite{Brazovskii_Muratov_1987_JETP,Kats_Lebedev_Muratov_1993_pr}, cholesteric liquid crystals~\cite{Brazovskii_Dmitriev_1975_JETP,Seul_Andelman_1995_science}, as well as some biological systems~\cite{Lavrentovich_Kamien_2016_pnas,Bates_Glinka_1988_prl}.
However, the symmetry of the order parameter in these systems implies a cubic term in the Landau free energy, and a first-order transition is generally not surprising.

In magnetic systems, Brazovskii scenario remains largely unstudied.
To our knowledge, the only other magnetic system where this type of transition has been clearly demonstrated is a helical magnet $\mathrm{MnSi}$~\cite{Muhlbauer_Boni_2009_Science,Janoschek_Boni_2013_prb,Bauer_Pfleiderer_2013_prl,Kindervater_Pfleiderer_2019_prx}.
Nevertheless, we believe that Brazovskii scenario applies to many other frustrated systems with large ground state degeneracy~\cite{Gvozdikova_zhitomirsky_2005_jetpl}. 
Our results demonstrate that the same arguments still hold for 2D systems with a 1D degenerate manifold, which is applicable to a large number of frustrated 2D magnets.

We note that higher-order effective interactions, which we ignored in this work, will break the degeneracy of the ring in $\chi_{0,D}(\mathbf{q})$.
However, provided that this splitting is small, thermal fluctuations will still populate the whole ring, meaning that the analysis above should still apply.

\textit{Concluding remarks.}-- Our theoretical study of the Ising-like phases in AB-SKL bilayers uncovered rich physical phenomena.
These phenomena bridge the properties of a broad range of magnetic and non-magnetic systems, such as liquid crystals, helical magnets, and glasses. 

The description of the ordered phases in our system depends crucially on the partial magnetic ordering of the unit cells, \textit{i.e.} the transition from Heisenberg to the XY chiral paramagnet.
The coarse-graining procedure that reflects this transition unveils the effective biquadratic interaction between LOPs, which is responsible for the stabilization of the chiral nematic phase.
To our knowledge, this work is the first to discuss a magnetic phase that is simultaneously nematic and chiral.
Even more remarkable is the fact that this phase is stabilized over a large range of parameters, thanks in part to the dual properties of the HDM model.
Furthermore, our analytical procedure can be generalized to other triangular systems, which could aid the experimental realization of chiral nematics.

We also note that our numerical results in the chiral paramagnet phase are qualitatively similar to the experimental results in the ``fluctuation disordered'' phase of $\mathrm{MnSi}$, appearing above $T_c$ at low fields~\cite{Pappas_Farago_2009_prl,Janoschek_Boni_2013_prb}.
Adding Brazovskii scenario to these similarities hints at universal properties of the HDM models that apply to crystals with and without inversion symmetry.

Finally, the glassy properties of the Ising-like phases deserve further investigation.
It is not clear if a \mbox{single-$q$} stripe always provides the ground state or if the Ising constraint may lead to additional frustration and, as a result, large degeneracy.
A combination of chirality and non-uniform magnetic structure makes these states an interesting subject for spintronic studies, since the itinerant electrons will couple to the emergent electromagnetic fields~\cite{Schulz_skyrmions_emergent_2012_nature,Everschor-Sitte_Sitte_the_2014_jap}.

To conclude, we hope that the richness of magnetic properties discussed in this work will serve as a motivation for future studies of the AB-SKL.

%%%%
%%%%
The work of A. Z., M. L. P., and T. L. M. was supported by the Natural Sciences and Engineering Research Council of Canada (NSERC).
The work of M. E. Z. was supported by ANR, France.

%

%\bibliography{references.bib}
\end{document}